\documentclass[11pt,a4paper]{article}

\usepackage[a4paper,margin=2.0cm]{geometry}
\usepackage{setspace}
\usepackage[utf8]{inputenc}
\usepackage[T1]{fontenc}
\usepackage{lmodern}
\usepackage{microtype}

\usepackage{graphicx}
\usepackage{xcolor}
\usepackage{colortbl}
\usepackage{amsmath}
\usepackage{amssymb}
\usepackage{siunitx}
\usepackage{steinmetz}
\usepackage{placeins}
\usepackage{authblk}
\usepackage[numbers,sort&compress]{natbib}
\usepackage[font=small,labelfont=bf]{caption}
\usepackage[colorlinks=true,
            linkcolor=blue,
            citecolor=blue,
            urlcolor=blue]{hyperref}

\graphicspath{{./images/}}

\title{\textbf{Boosting self-hybridized exciton polaritons with metal-clad WS$_2$ waveguides}}

\author[1]{Filip Majstorovic}
\author[1]{Masoud Taleb}
\author[1]{Victor DeManuel-Gonzalez}
\author[1,2]{Kai Rossnagel}
\author[1]{Nahid Talebi}

\affil[1]{Institute of Experimental and Applied Physics, Kiel University, 24098 Kiel, Germany}
\affil[2]{Ruprecht Haensel Laboratory, Deutsches Elektronen-Synchrotron DESY, 22607 Hamburg, Germany}

\date{}

\begin{document}

\maketitle

\begin{abstract}
The formation of Fabry--P\'erot and guided-wave self-hybridized exciton-polaritons in two-dimensional materials results in long-range exciton energy transfer and strong exciton--exciton interactions. Here, we demonstrate that the coupling strength between photonic modes and excitons is significantly boosted by embedding the active excitonic layer in a metal-clad WS$_2$ waveguide. The photonic modes in this waveguide exhibit modified dispersion properties for both Fabry--P\'erot-type and guided-wave exciton-polaritons compared to pure WS$_2$ flakes, and show an increased coupling strength. Our results provide a robust approach for controlling exciton--photon interactions and their coupling strength in hybrid heterostructures.
\end{abstract}

\vspace{0.5em}

\vspace{0.5em}
\noindent\textbf{Correspondence:} Filip Majstorovic (\href{mailto:majstorovic@physik.uni-kiel.de}{majstorovic@physik.uni-kiel.de}) and Nahid Talebi (\href{mailto:talebi@physik.uni-kiel.de}{talebi@physik.uni-kiel.de})

\vspace{1em}

\section{Introduction}
\label{sec:introduction}

Light in photonic and plasmonic cavities coupled to excitons or other material excitations, such as quantum dots or molecules, can host hybrid light--matter states known as polaritons. These states arise in the strong-coupling regime, where the interaction between the optical mode and the material excitation leads to the formation of new eigenstates \cite{Zhen2022,Chikkaraddy2016,Yankovich2019,strong_coupling_plasmonic_particle_and_QD,Zhao2025}.
The strong-coupling regime is reached when the energy exchange rate between the light and matter components exceeds their respective damping rates \cite{You2023,Luo2024,Trm2015}.
When light is strongly coupled specifically to excitons, the resulting hybrid quasiparticles are referred to as exciton-polaritons. These states have attracted considerable interest due to their unique physical properties and their potential for applications such as Bose--Einstein condensation, polariton lasing, solar energy harvesting, superfluidity and potentially quantum computing \cite{Deng2010_Bose_Einstein_Condensation,Zhang2022_Polariton_Lasers,Harder2021_Topological_Polariton_Laser,Nikolis2019_EP_Photovoltaics,Peng2022_EP_Superfluidity,Kavokin2022_polariton_quantum_computing}.
Semiconducting transition metal dichalcogenides (TMDCs), such as WS$_2$, are particularly attractive materials for providing room-temperature excitons with properties well suited for reaching the strong-coupling regime. Monolayer WS$_2$ crystals, in contrast to the bulk material, exhibit a direct band gap. Owing to their reduced dimensionality, few-layer systems also exhibit reduced dielectric screening. As a result, they host excitons with large oscillator strengths and high binding energies of up to several hundred meV \cite{Luo2024,Wang2018_Excitons_in_atomically_thin_TMDCs,Chernikov2014,Chahshouri2022}.
It has already been shown that thin films of TMDCs can themselves support guided optical modes and thus provide both the excitonic and the photonic components within the same material system \cite{Khurgin2015,Fei2016,Munkhbat2019_Self_hybridized_EP,Taleb2022_Charting_the_EP_landscape_of_WSe2_CL}. As a result, they can host exciton-polaritons even without an external resonator, giving rise to so-called self-hybridized exciton-polaritons \cite{Munkhbat2019_Self_hybridized_EP,Taleb2022_Charting_the_EP_landscape_of_WSe2_CL,Shin2022,Zong2023}. These hybrid modes can enable long-range exciton energy transport in TMDC crystals \cite{Taleb2022_Charting_the_EP_landscape_of_WSe2_CL}.
However, an external cavity, most simply realized by two parallel mirrors, can improve the quality of the optical mode and therefore enhance the interaction between the photon and the exciton \cite{Weisbuch1992,Dufferwiel2015_EP_Photonic_cavity,Flatten2016}. Exciton-polaritons have already been realized in a variety of externally defined photonic and plasmonic structures, including dielectric microcavities, plasmonic cavities, and confined dielectric or plasmonic waveguides \cite{Dufferwiel2015_EP_Photonic_cavity,Flatten2016,Wu2022,Kleemann2017_EP_plasmonic_cavity,Dibos2019}. Plasmonic cavities and waveguides are particularly attractive because they provide strong electromagnetic field confinement and can support surface plasmon polaritons (SPPs) or, depending on the geometry, localized plasmon resonances \cite{Chikkaraddy2016,Trm2015,Sun2021,Davoodi2021}. Besides dielectric cavity and guided modes, plasmonic modes can also strongly interact with excitons, giving rise to plasmon-exciton-polaritons \cite{Kleemann2017_EP_plasmonic_cavity,Dibos2019,Sun2021,Davoodi2021,Liu2016,Wang2019,Zhang2024_WS2_EP_Plasmon_coupling}.
Related planar metal--TMDC geometries have been investigated as compact platforms for plasmon-exciton-polaritons and nanoscale light--matter interaction \cite{Davoodi2021,Gonalves2018}. 

In this work, we theoretically and experimentally investigate a planar Au/WS$_2$/Au heterostructure using cathodoluminescence spectroscopy. The structure consists of a few-layer WS$_2$ crystal embedded between two thin gold layers, which act as partially reflecting mirrors and form a metal-clad WS$_2$ waveguide. The WS$_2$ layer therefore provides both the excitonic resonance and the guiding medium, while the metallic boundaries enhance the optical field confinement. By solving Maxwell's equations for this layered geometry, we calculate the allowed optical modes and their dispersion and compare them with the cathodoluminescence measurements.
This structure can support guided modes with Fabry--P\'erot-type resonances and also surface plasmon polaritons at the Au/air or Au/WS$_2$ interfaces \cite{Trm2015,Davoodi2021,Zhang2024_WS2_EP_Plasmon_coupling,Gonalves2018}. We show that all these optical modes can strongly interact with excitons when they are spectrally resonant, have sufficient spatial field overlap with the excitonic material, while their coupling rate exceeds the relevant loss rates \cite{Luo2024,Trm2015,Sun2021}. In the present work, however, we focus on the strong coupling between the supported guided modes and the excitons provided by the WS$_2$ layer.
We show that the gold cladding significantly enhances the coupling strength $G=\hbar g$ between these guided modes and the excitons, reaching $G=\SI{190}{\milli\electronvolt}$. This value is larger than values reported for related self-hybridized TMDC systems, including WSe$_2$ thin flakes with $G=\SI{120}{\milli\electronvolt}$ \cite{Taleb2022_Charting_the_EP_landscape_of_WSe2_CL} and WS$_2$ nanotube waveguides with $G=\SI{163}{\milli\electronvolt}$ \cite{Yadgarov2018}.

\section{Results and Discussion}
\label{sec:results_discussion}

\subsection{Sample characterization and cathodoluminescence spectroscopy}
\label{subsec:sample_characterization_cl}

The investigated sample consists of several Au/WS$_2$/Au heterostructures fabricated on a silica glass substrate. Each structure is defined by two approximately $\SI{50}{\nano\meter}$ thick circular gold layers, between which WS$_2$ flakes are embedded. The WS$_2$ flakes studied in this work have thicknesses ranging from about $\SI{90}{\nano\meter}$ to $\SI{130}{\nano\meter}$.
For the fabrication, twelve circular bottom gold layers with a diameter of $\SI{200}{\micro\meter}$ were first deposited onto the substrate. WS$_2$ crystals were then mechanically exfoliated onto the gold-patterned areas, resulting in multiple flakes on each circle. Subsequently, twelve circular top gold layers with the same diameter were deposited onto the flakes. The regions in which WS$_2$ flakes overlap with both the bottom and the top gold layer therefore form individual Au/WS$_2$/Au heterostructures.
These structures were experimentally investigated using spatially and angle-resolved cathodoluminescence (CL) spectroscopy in a scanning electron microscope (SEM) at acceleration voltages of $\SI{10}{\kilo\electronvolt}$, $\SI{15}{\kilo\electronvolt}$, and $\SI{20}{\kilo\electronvolt}$. In CL spectroscopy, the focused electron beam acts as a local broadband excitation source, while the emitted radiation is collected and spectrally analyzed \cite{Talebi2017}. 

Owing to the near-field character of the electron-induced excitation, CL can access optical modes over a wide range of in-plane momenta positioned within the light cone, including guided modes and surface plasmon polaritons \cite{GarciaDeAbajo2010,Bittorf2026_long_range_exciton_energy_transfer}.
The emitted light was collected with a parabolic mirror placed above the sample and directed to a spectrometer. 

\begin{figure}[htb]
\centering
\includegraphics[width=\linewidth, trim=0cm 8cm 0cm 0cm, clip]{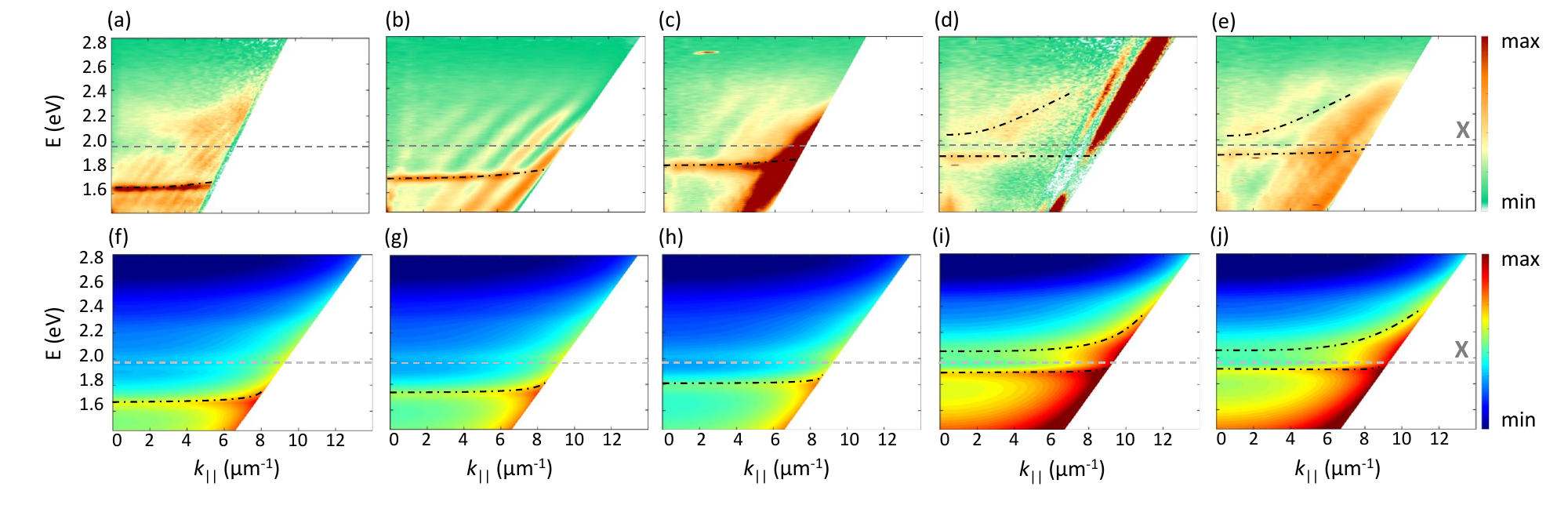}
\caption{Momentum-resolved cathodoluminescence spectra of Au/WS$_2$/Au samples. The WS$_2$ thicknesses in (a) and (b) were independently determined by AFM to be $\SI{130}{\nano\meter}$ and $\SI{120}{\nano\meter}$, while the thicknesses in (c)--(e) were estimated by comparison with the calculated dispersion relations shown in (f)--(j). The grey dashed line indicates the A-exciton energy at $\SI{1.97}{\electronvolt}$. (f)--(j) Calculated dispersion relations of an idealized Au/WS$_2$/Au system for WS$_2$ thicknesses of $\SI{130}{\nano\meter}$, $\SI{120}{\nano\meter}$, $\SI{110}{\nano\meter}$, $\SI{95}{\nano\meter}$, and $\SI{90}{\nano\meter}$, respectively. The calculated dispersions in (i) and (j), together with the corresponding experimental maps in (d) and (e), show a clear anticrossing near the A-exciton energy.}
\label{fig:Figure_1}
\end{figure}

Figures~\ref{fig:Figure_1}(a)--(e) present momentum-resolved CL spectra acquired from five different WS$_2$ flakes with varying thicknesses. The corresponding calculated dispersion relations are arranged directly below the experimental maps in Figures~\ref{fig:Figure_1}(f)--(j) and were calculated for WS$_2$ thicknesses of $\SI{130}{\nano\meter}$, $\SI{120}{\nano\meter}$, $\SI{110}{\nano\meter}$, $\SI{95}{\nano\meter}$, and $\SI{90}{\nano\meter}$, respectively. The dispersions were obtained by solving Maxwell's equations for the layered geometry following standard multilayer optics approaches \cite{Talebi2019_book}. The frequency-dependent relative permittivities of gold and WS$_2$ were used as material input and calculated from the tabulated refractive index $n$ and extinction coefficient $k$ according to $\varepsilon_r(\omega)=[n(\omega)+\mathrm{i}k(\omega)]^2$. The optical constants of gold were taken from Yakubovsky \textit{et al.} \cite{Yakubovsky2017}, while those of WS$_2$ were taken from Vyshnevyy \textit{et al.} \cite{Vyshnevyy2023}. 

For the samples in Figures~\ref{fig:Figure_1}(a) and (b), the flake thicknesses were independently determined by atomic force microscopy (AFM) to be approximately $\SI{130}{\nano\meter}$ and $\SI{120}{\nano\meter}$. The remaining thicknesses were inferred from the agreement between the measured and calculated dispersions.

In all dispersion maps, the A-exciton energy at $\SI{1.97}{\electronvolt}$ is indicated by a grey dashed line. Since the CL emission is detected in the far field, the experimental maps show the radiatively outcoupled part of the dispersion within the light cone.
Comparing the angle-resolved CL maps in Figures~\ref{fig:Figure_1}(a)--(e) with the calculated dispersion relations in Figures~\ref{fig:Figure_1}(f)--(j), it can be seen that the experimental maps do not cover the full area inside the light cone. In addition, the experimentally accessible $E$--$k$ range varies between the different measurements. This is caused by the finite collection angle of the parabolic mirror used in the CL setup. Since the numerical aperture of the mirror is smaller than unity, not all emission angles inside the light cone are collected. Furthermore, the collection geometry is not perfectly symmetric. The parabolic mirror contains an aperture for the electron beam, and the mirror shape around this aperture leads to a different angular collection range on opposite sides \cite{Coenen2017}. As a result, CL maps recorded from different sides of the mirror aperture can cover slightly different regions in momentum space. This explains why the experimental maps in Figures~\ref{fig:Figure_1}(a)--(e) show different portions of the light cone, while the calculated dispersions in Figures~\ref{fig:Figure_1}(f)--(j) display the complete light-cone region.

All momentum-resolved CL spectra exhibit a pronounced nearly horizontal resonance below the A-exciton energy. The energy of this resonance increases with decreasing WS$_2$ thickness, indicating that it originates from a thickness-dependent optical mode of the Au/WS$_2$/Au waveguide. Comparison with the calculated dispersion relations shown in Figures~\ref{fig:Figure_1}(a)--(f) and Figure~\ref{fig:Figure_5}(a) identifies this resonance as the radiatively accessible part of the guided mode in the heterostructure.
For the thinner flakes in Figures~\ref{fig:Figure_1}(d) and (e), the mode approaches the A-exciton energy and shows a clear anticrossing behavior. At the same time, an additional resonance appears above the exciton energy. These two resonances can therefore be assigned to the lower and upper polariton branches. The calculated dispersion relations in Figures~\ref{fig:Figure_1}(i) and (j) reproduce this behavior and further support the assignment of the observed modes to strongly coupled guided exciton-polariton branches. In addition, Figures~\ref{fig:Figure_1}(a) and (b) exhibit pronounced vertical interference fringes. These fringes are likely caused by interference between transition radiation (TR) and surface plasmon polaritons (SPPs), which propagate towards the flake edges and radiate from there. This interpretation will be supported by a quantitative fit later below.  

\begin{figure}[t]
\centering
\includegraphics[width=0.85\textwidth, trim=0cm 2cm 11cm 0cm, clip]{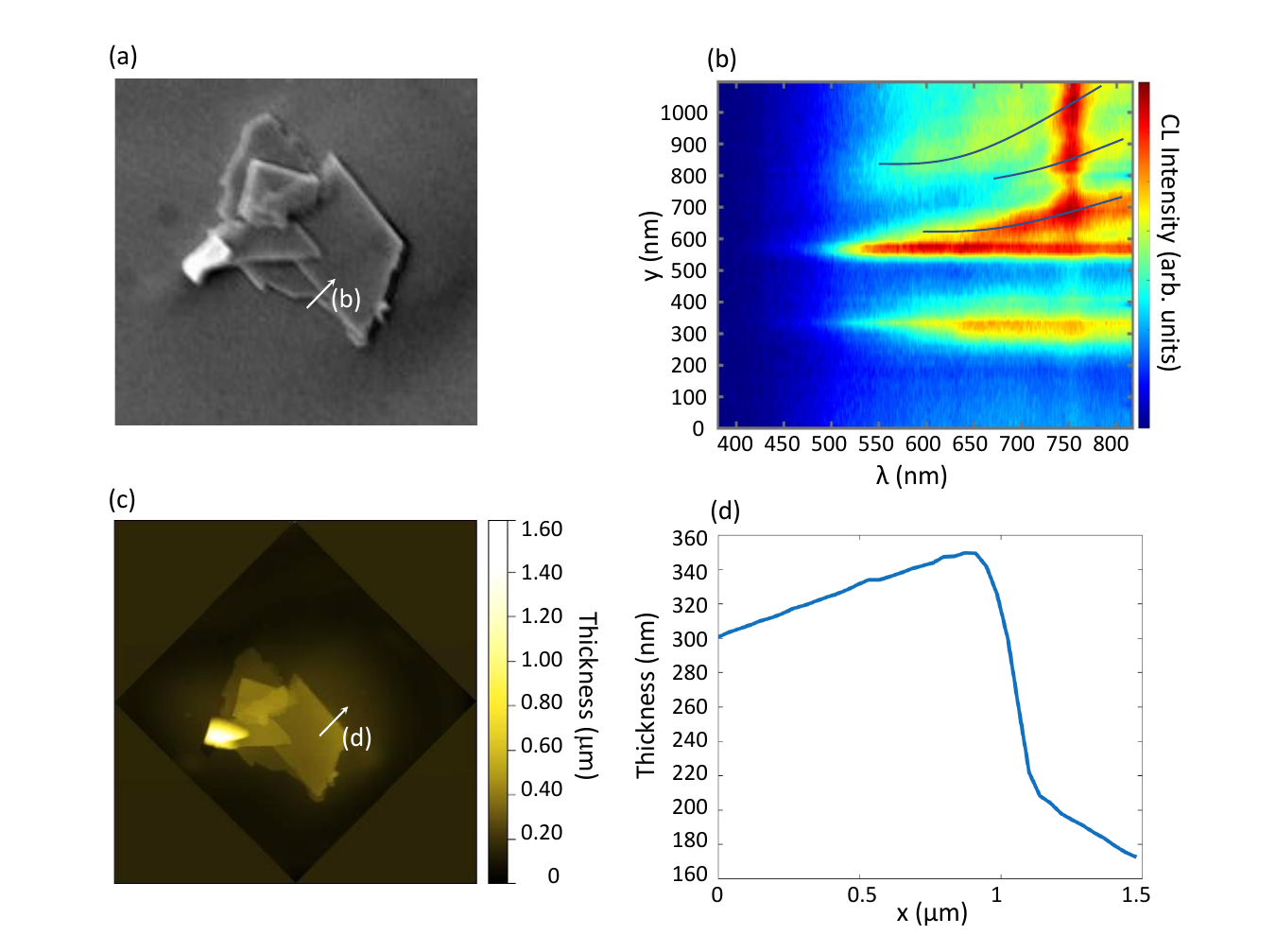}
\caption{(a) SEM image of the WS$_2$ flake with a thickness of approximately $\SI{130}{\nano\meter}$. (b) Spatially resolved CL map acquired over a distance of $\SI{1}{\micro\meter}$ along the path indicated by the white arrow in (a). A pronounced resonance is observed around $\SI{750}{\nano\meter}$, which is associated with the lower polariton branch of the guided mode identified in the momentum-resolved CL spectrum in Figure~\ref{fig:Figure_1}(a). (c) AFM image of the same flake, where the color scale represents the sample height. The white arrow marks the position of the height profile shown in (d). (d) One-dimensional height profile across the flake, yielding a thickness of approximately $\SI{130}{\nano\meter}$. The slope in the height profile is attributed to a bubble formed in the gold film during electron-beam irradiation.}
\label{fig:Figure_2}
\end{figure}

Figure~\ref{fig:Figure_2}(a) presents an SEM image of the WS$_2$ flake with a thickness of approximately $\SI{130}{\nano\meter}$, corresponding to the flake investigated in Figure~\ref{fig:Figure_1}(a). The white arrow marks the line along which the spatially resolved CL spectrum in Figure~\ref{fig:Figure_2}(b) was recorded. In this spectrum, the vertical axis corresponds to the position along the line scan, the horizontal axis to the emission wavelength, and the color scale to the CL intensity.
The spatially resolved CL map exhibits a pronounced resonance at a wavelength of about $\SI{750}{\nano\meter}$, which corresponds to the resonance observed in the angle-resolved CL spectrum in Figure~\ref{fig:Figure_1}(a) at approximately $\SI{1.65}{\electronvolt}$. The angle-resolved spectrum shows that this resonance appears nearly dispersionless over the experimentally accessible momentum range. This indicates a strongly reduced group velocity $v_\mathrm{g}=\mathrm{d}\omega/\mathrm{d}k_{||}=(1/\hbar)\,\mathrm{d}E/\mathrm{d}k_{||}$ of the guided lower-polariton mode.
Such a small group velocity is interesting because slow optical modes can increase the effective interaction time between the optical mode and the excitonic medium and enhance the photonic density of states, thereby strengthening effective light--matter interaction in waveguide systems \cite{Baba2008,Moerk2010}. In polaritonic systems, flat or weakly dispersive modes are also relevant for collective phenomena such as polariton condensation, since they can favor mode occupation and relaxation into selected low-energy states \cite{Deng2010}.

In addition, the spectrum displays interference fringes with three pronounced intensity maxima. Similar to the vertical features observed in the angle-resolved CL spectra in Figures~\ref{fig:Figure_1}(a) and (b), these fringes are attributed to the interference between transition radiation and surface plasmon polaritons.
Figures~\ref{fig:Figure_2}(c) and (d) provide the AFM characterization of the same flake acquired after CL measurements. Figure~\ref{fig:Figure_2}(c) displays the two-dimensional height map, while Figure~\ref{fig:Figure_2}(d) displays the height profile along the line indicated by the white arrow in Figure~\ref{fig:Figure_2}(c). The pronounced step in this profile corresponds to the height difference between the surrounding surface and the upper surface of the WS$_2$ flake. From this step height, the thickness of the investigated flake is determined to be approximately $\SI{130}{\nano\meter}$. After extensive CL measurements and due to charges deposited into the glass substrate, the heterostructure is formed into a bubble shape, visible from the nonflat AFM map.

\begin{figure}[t]
\centering
\includegraphics[width=0.95\textwidth, trim=0cm 9cm 1cm 0cm, clip]{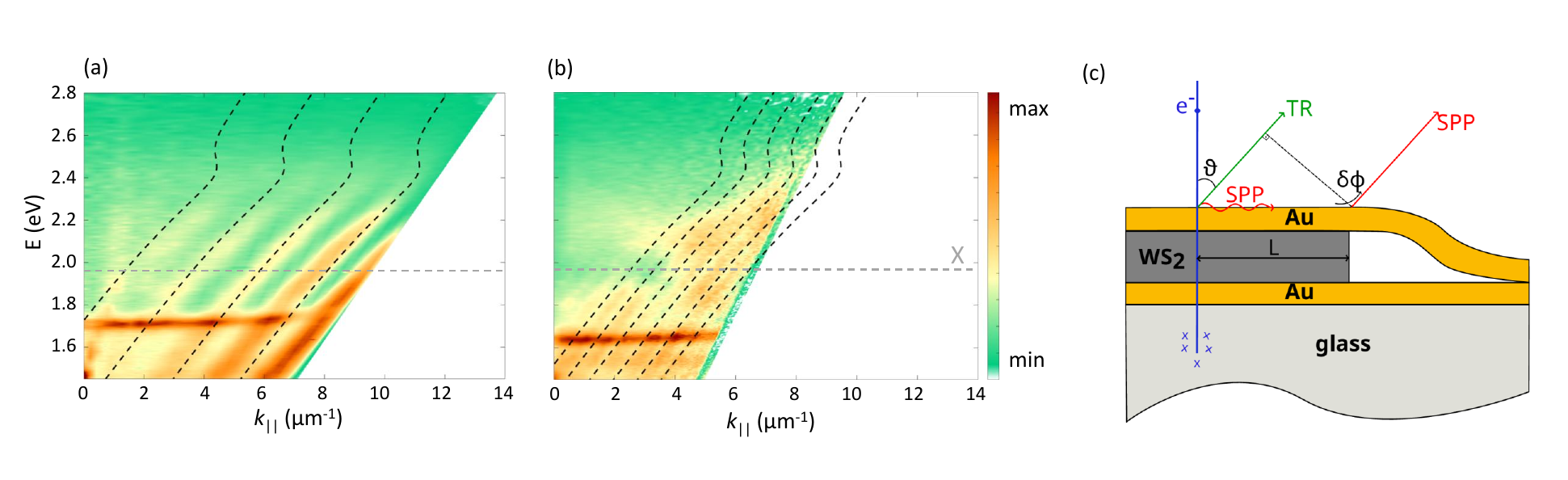}
\caption{Experimental angle-resolved cathodoluminescence spectra of Au/WS$_2$/Au flakes with WS$_2$ thicknesses of (a) $\SI{120}{\nano\meter}$ and (b) $\SI{130}{\nano\meter}$, corresponding to the samples already shown in Figure~\ref{fig:Figure_1}. The maps are overlaid with fits to calculated branches arising from the interference of surface plasmon polaritons (SPPs) and transition radiation (TR). The grey dashed line indicates the A-exciton energy at $\SI{1.97}{\electronvolt}$. (c) Schematic side view of the sample and of the interference mechanism: SPPs are excited at the electron-beam impact position, propagate towards the flake edge, are scattered there into the far field, and interfere with transition radiation.}
\label{fig:Figure_3}
\end{figure}

The vertical interference fringes observed in the angle-resolved CL maps in Figures~\ref{fig:Figure_1}(a)--(e) can be explained by the interference between TR and SPPs. To verify this interpretation, the experimental CL spectra were compared with the corresponding interference condition. The underlying process is schematically illustrated in Figure~\ref{fig:Figure_3}(c). The incident electron beam generates TR directly at the excitation position and simultaneously excites SPPs. These SPPs propagate along the gold layer towards the WS$_2$ flake edge, where they are scattered into the far field. The far-field radiation originating from the scattered SPPs can then interfere with the directly emitted TR.
The resulting interference maxima follow the condition \cite{Guo2019_Far_Field_Radiation_3D_Plasmonic_Au_Tapers}
\begin{equation}
    L_{\mathrm{eff}} k_{||}
    = n_{\mathrm{eff}}(\omega) \frac{\omega}{c} L_{\mathrm{eff}} + 2\pi m .    
\label{eq:TR_SPP_interference}
\end{equation}
In this expression, $L_{\mathrm{eff}}$ denotes the effective propagation distance between the electron-beam excitation position and the scattering edge of the flake. The in-plane wave vector of the detected radiation is denoted by $k_{||}$, while $n_{\mathrm{eff}}(\omega)$ is the frequency-dependent effective refractive index of the SPP mode given by the ratio between the phase constant of SPP and free-space wave number of the light. Furthermore, $\omega$ is the angular frequency, $c$ is the speed of light in vacuum, and $m$ is an integer. Phase shifts accumulated during SPP excitation, propagation, and scattering from the edge are absorbed into the effective propagation length $L_{\mathrm{eff}}$.
Figures~\ref{fig:Figure_3}(a) and (b) show the angle-resolved CL spectra from Figures~\ref{fig:Figure_1}(a) and (b), overlaid with the calculated interference maxima from Equation~\eqref{eq:TR_SPP_interference} for different interference orders $m$. The best agreement is obtained for effective propagation distances of $L_{\mathrm{eff}}=\SI{2.8}{\micro\meter}$ and $L_{\mathrm{eff}}=\SI{8}{\micro\meter}$ for Figures~\ref{fig:Figure_3}(a) and (b), respectively. The values of $L_{\mathrm{eff}}$ were chosen to reproduce the experimentally observed fringe positions. These values are on a similar length scale as previous CL interference experiments. For example, Schilder \textit{et al.} reported TR--SPP interference for electron-beam positions about $\SI{2.3}{\micro\meter}$ to $\SI{2.4}{\micro\meter}$ away from plasmonic scatterers in a silver film on a SiO$_x$ substrate \cite{Schilder2020_L_value}, treating this distance as the geometrical propagation length and including additional phase shifts separately. Kuttge \textit{et al.} demonstrated far-field interference between TR and SPPs on gold gratings over micrometer-scale beam--grating distances \cite{Kuttge2009_L_value}.
The calculated branches reproduce the experimentally observed vertical fringes well. In particular, the curvature of several fringes around $\SI{2.4}{\electronvolt}$ is captured, most clearly in Figure~\ref{fig:Figure_3}(a). In addition, the fringes disappear for energies above approximately $\SI{2.5}{\electronvolt}$, consistent with the spectral range in which SPP-mediated interference is expected to contribute for gold. This further supports the assignment of the vertical lines to the interference between TR and SPP-mediated radiation from the edge.

\subsection{Mode identification and coupling-strength analysis}
\label{subsec:mode_identification_coupling}

\begin{figure}[t]
\centering
\includegraphics[width=0.95\textwidth, trim=0cm 4cm 0cm 0cm, clip]{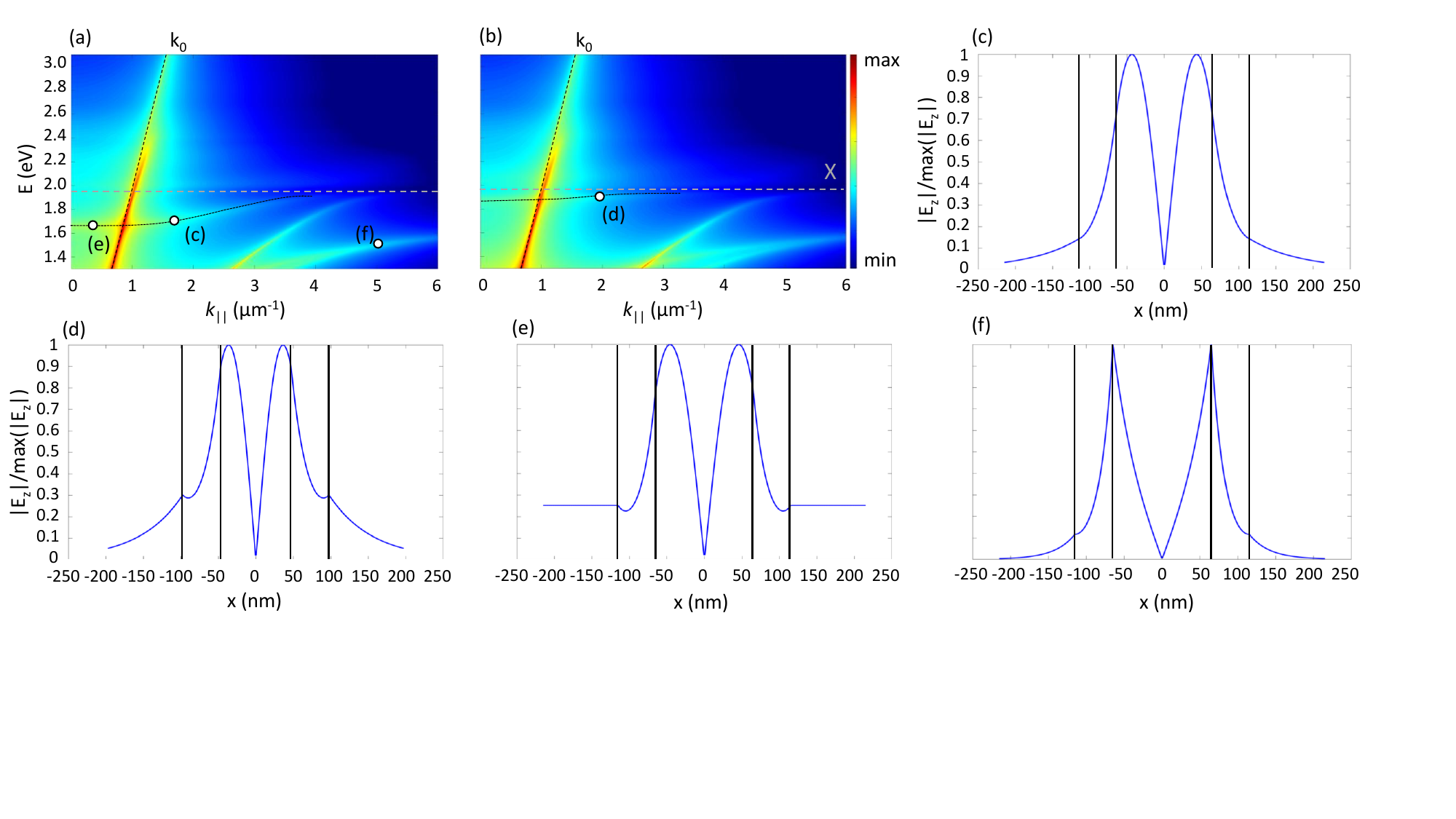}
\caption{(a,b) Calculated dispersion relations of idealized Au/WS$_2$/Au structures with WS$_2$ thicknesses of $\SI{130}{\nano\meter}$ and $\SI{95}{\nano\meter}$, including the region outside the light cone. The white dots mark the positions at which the corresponding field profiles shown in (c)--(f) were calculated. The grey dashed line indicates the A-exciton energy at $\SI{1.97}{\electronvolt}$. (c)--(f) Normalized electric field profiles, $|E_z|/\max(|E_z|)$, for the modes marked in (a) and (b). The black vertical lines indicate the interfaces between air, gold, and WS$_2$. Panels (c) and (d) correspond to the lower polariton outside the light cone, while (e) shows the lower polariton inside the light cone, where the field is no longer evanescent in the outer regions due to radiative outcoupling. In contrast to the SPP mode, the lower-polariton field exhibits an oscillatory profile within the WS$_2$ layer. Panel (f) shows the field profile of the SPP mode, where the field is confined to the WS$_2$/Au interfaces and decays evanescently away from them on both sides.}
\label{fig:Figure_4}
\end{figure}

To further analyze the nature of the calculated modes, we consider the full dispersion relation of the Au/WS$_2$/Au structure, including both the region inside and outside the light cone. These calculations follow the same multilayer optics approach described above, but are shown over an extended momentum range. Based on the same calculation, the corresponding electric-field profiles of the supported modes were calculated at selected points along the dispersion.
Figures~\ref{fig:Figure_4}(a) and (b) present the calculated dispersion relations for WS$_2$ thicknesses of $\SI{130}{\nano\meter}$ and $\SI{95}{\nano\meter}$, respectively. In contrast to the calculated maps shown in Figures~\ref{fig:Figure_1}(f)--(j), the full momentum range is displayed here, including modes outside the light cone, which are out of the reach of the CL technique. For the $\SI{130}{\nano\meter}$ thick flake in Figure~\ref{fig:Figure_4}(a), the mode observed at approximately $\SI{1.65}{\electronvolt}$ inside the light cone continues to larger in-plane momenta outside the light cone. There, it approaches the WS$_2$ A-exciton energy and exhibits a pronounced anticrossing, identifying it as the lower polariton branch of a strongly coupled guided mode. For the $\SI{95}{\nano\meter}$ thick flake in Figure~\ref{fig:Figure_4}(b), the anticrossing already becomes visible inside the light cone, while the mode continues further outside the light cone at larger momenta.

In addition to the guided polariton branches, the calculated dispersions contain surface plasmon polaritons. One SPP mode appears at larger in-plane momenta and is split into two branches due to the two Au/WS$_2$ interfaces of the heterostructure.
Another SPP mode appears close to the light line.
Figures~\ref{fig:Figure_4}(c)--(f) show the normalized $z$-component of the electric field, $|E_z|/\max(|E_z|)$, across the Au/WS$_2$/Au cross section.
Figures~\ref{fig:Figure_4}(c) and (d) correspond to the lower polariton branch outside the light cone for the $\SI{130}{\nano\meter}$ and $\SI{95}{\nano\meter}$ thick flakes, respectively. In both cases, the field exhibits an oscillatory profile inside the WS$_2$ layer and decays evanescently outside the structure. This field distribution is characteristic of a guided mode confined mainly within the WS$_2$ layer.
Figure~\ref{fig:Figure_4}(e) shows the field profile of the same lower polariton branch for the $\SI{130}{\nano\meter}$ thick flake, but at a momentum inside the light cone. The field still oscillates inside the WS$_2$ layer, but it no longer decays evanescently in the outer regions. Instead, the field extends into the surrounding medium, reflecting the radiative character of modes inside the light cone. Figure~\ref{fig:Figure_4}(f) displays the field profile of the surface plasmon polariton. Here, the field is strongly localized at the Au/WS$_2$ interfaces and decays evanescently away from them. This clearly distinguishes the SPP mode from the guided lower polariton, whose field is distributed across the WS$_2$ layer.

\begin{figure}[t]
\centering
\includegraphics[width=0.95\textwidth, trim=0cm 4cm 0cm 3cm, clip]{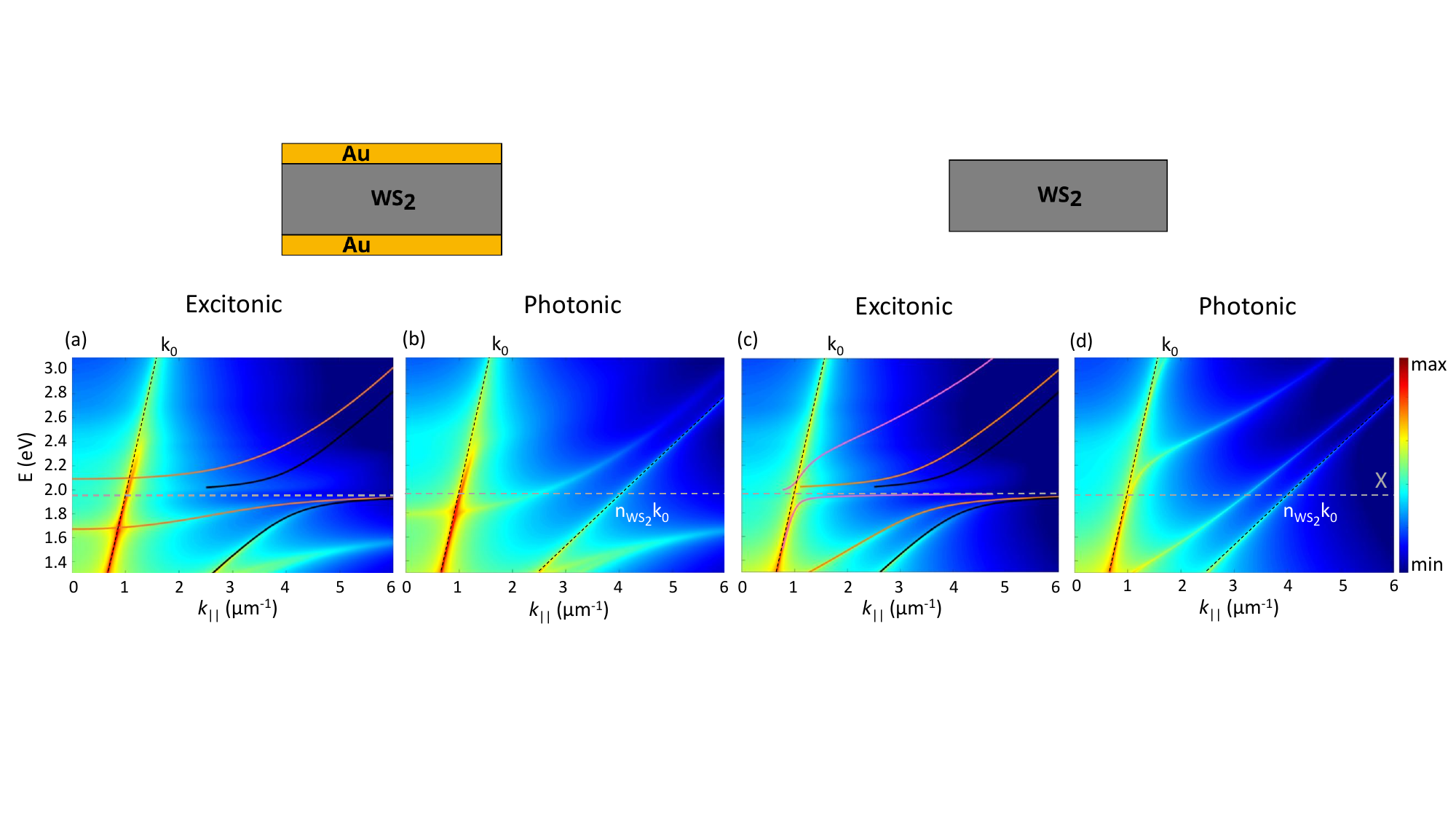}
\caption{(a) Calculated dispersion relation of the Au/WS$_2$/Au structure with a WS$_2$ thickness of $\SI{130}{\nano\meter}$ and a gold thickness of $\SI{50}{\nano\meter}$, including the region outside the light cone. Hopfield fits to the lower and upper polariton branches of the guided mode and of the bent light line in WS$_2$ are overlaid. For both the black and orange branches, the coupling strength is $\SI{190}{\milli\electronvolt}$. (b) Dispersion relation of the same structure using the constant-plus-Drude background permittivity of WS$_2$ instead of the full WS$_2$ dielectric function, so that the excitonic resonances and the UV oscillator are removed and only the optical modes remain visible. The gold thickness is again $\SI{50}{\nano\meter}$. (c) Dispersion relation for the same WS$_2$ thickness, but with a gold thickness of $\SI{1}{\nano\meter}$ in order to suppress the influence of the gold layers. Hopfield fits to the lower and upper polariton branches are shown, and an additional mode appears. The corresponding coupling strengths for the black, orange, and pink branches are $\SI{190}{\milli\electronvolt}$, $\SI{180}{\milli\electronvolt}$, and $\SI{120}{\milli\electronvolt}$, respectively. (d) Dispersion relation for the same structure using again the constant-plus-Drude background permittivity of WS$_2$. In this case, the guided modes remain between the air and WS$_2$ light lines, since the gold layers are effectively absent, and the SPP modes disappear. The grey dashed line marks the A-exciton energy at $\SI{1.97}{\electronvolt}$.}
\label{fig:Figure_5}
\end{figure}

After identifying the relevant guided modes and their electric-field distributions, we now analyze their coupling to the WS$_2$ excitons more quantitatively. The formation of polariton modes can be described using the Hopfield model, in which the photonic mode and the excitonic transition are treated as coupled quantum-mechanical oscillators \cite{Hopfield1958}.
For a single optical mode interacting with an excitonic transition described as a two-level system, the Hopfield Hamiltonian can be written as
\begin{equation}
\hat{H} =
\hbar \omega_c \hat{a}^\dagger \hat{a}
+ \hbar \omega_b \hat{b}^\dagger \hat{b}
+ \mathrm{i}\hbar g
\left(\hat{a}\hat{b}^\dagger - \hat{a}^\dagger\hat{b}\right),
\label{eq:Hopfield_Hamiltonian}
\end{equation}
within the rotating wave approximation.
The frequencies of the optical mode and the exciton are denoted by $\omega_c$ and $\omega_b$, respectively. The operators $\hat{a}^\dagger$ and $\hat{a}$ create and annihilate a photon, while $\hat{b}^\dagger$ and $\hat{b}$ create and annihilate an exciton, respectively. 

The lower and upper polariton energies are then obtained from the eigenvalues of the Hamiltonian in Equation~\eqref{eq:Hopfield_Hamiltonian}, yielding
\begin{equation}
E_{\mathrm{LP,UP}} (k_{||})
=
\frac{1}{2}\left(E_b + E_c (k_{||}) \right)
\pm
\frac{1}{2}
\sqrt{\left(E_b - E_c (k_{||}) \right)^2 + 4G^2}.
\label{eq:polariton_energies}
\end{equation}
In this expression, $E_b=\hbar\omega_b$ and $E_c=\hbar\omega_c$ are the exciton and optical-mode energies, respectively, while $G=\hbar g$ is the coupling strength expressed as an energy.

Equation~\eqref{eq:polariton_energies} is fitted to the calculated lower and upper polariton branches in Figures~\ref{fig:Figure_5}(a) and (c). In both cases, the WS$_2$ thickness is $\SI{130}{\nano\meter}$. Figure~\ref{fig:Figure_5}(a) corresponds to the full Au/WS$_2$/Au structure with two $\SI{50}{\nano\meter}$ thick gold layers, while Figure~\ref{fig:Figure_5}(c) shows the corresponding calculation for gold layers with a thickness of only $\SI{1}{\nano\meter}$. The latter case is used to reduce the influence of the gold cladding and to approximate the behavior of an almost uncladded WS$_2$ waveguide.
To apply Equation~\eqref{eq:polariton_energies}, the dispersion of the uncoupled optical-mode energy $E_c$ has to be known. Therefore, we additionally calculate the dispersion relations after removing the excitonic resonances from the WS$_2$ permittivity. These calculations are shown in Figures~\ref{fig:Figure_5}(b) and (d) for gold thicknesses of $\SI{50}{\nano\meter}$ and $\SI{1}{\nano\meter}$, respectively. In these calculations, the full dispersive WS$_2$ permittivity is replaced by a smooth background permittivity, so that only the optical modes remain visible.

The background permittivity is obtained by fitting the full dispersive dielectric function $\varepsilon_r(\omega)$ of WS$_2$ with a Drude--Lorentz model. The model contains a constant high-energy background, a Drude contribution, Lorentz oscillators for the A-, B-, and C-excitonic resonances, and an additional Lorentz oscillator for the higher-energy UV background, corresponding to the WS$_2$ above-band-gap absorption. The full model for the WS$_2$ permittivity is therefore given by
\begin{equation}
\begin{aligned}
\varepsilon_{\mathrm{model}}(E)
={}& \varepsilon_\infty
- \frac{E_\mathrm{p}^2}
{E\left(E+\mathrm{i}\gamma_\mathrm{D}\right)}
+ \sum_{j=\mathrm{A},\mathrm{B},\mathrm{C}}
\frac{A_j E_{\mathrm{ex},j}^2}
{E_{\mathrm{ex},j}^2 - E^2 - \mathrm{i}\gamma_{\mathrm{ex},j}E} \\
&+ \frac{A_\mathrm{UV} E_{\mathrm{UV}}^2}
{E_{\mathrm{UV}}^2 - E^2 - \mathrm{i}\gamma_{\mathrm{UV}}E}.
\end{aligned}
\label{eq:drude_lorentz_model}
\end{equation}
In Equation~\eqref{eq:drude_lorentz_model}, $\varepsilon_\infty$ denotes the constant high-energy background permittivity. The parameters $E_\mathrm{p}$ and $\gamma_\mathrm{D}$ describe the plasma energy and damping of the Drude contribution. The parameters $A_j$, $E_{\mathrm{ex},j}$, and $\gamma_{\mathrm{ex},j}$ correspond to the oscillator strength, resonance energy, and damping of the A-, B-, and C-excitons, respectively. The parameters $A_\mathrm{UV}$, $E_{\mathrm{UV}}$, and $\gamma_{\mathrm{UV}}$ denote the corresponding oscillator strength, resonance energy, and damping of the UV background oscillator. The corresponding fit parameters are provided in Table~\ref{tab:DL_fit_parameters}.
For the background calculations in Figures~\ref{fig:Figure_5}(b) and (d), only the constant term and the Drude contribution are retained,
\begin{equation}
\varepsilon_{\mathrm{bg}}(E)
=
\varepsilon_\infty
-
\frac{E_\mathrm{p}^2}
{E\left(E+\mathrm{i}\gamma_\mathrm{D}\right)} .
\label{eq:drude_background}
\end{equation}
This removes the A-, B-, and C-excitonic resonances as well as the UV oscillator, so that the uncoupled optical modes of the structure can be identified.

\begin{table}[t]
\centering
\caption{Fit parameters of the Drude--Lorentz model used for the WS$_2$ permittivity.}
\label{tab:DL_fit_parameters}
\begin{tabular}{llll}
\hline
Parameter & Description & Value & Unit \\
\hline
$\varepsilon_\infty$ & constant background permittivity & 8 & -- \\
$E_\mathrm{p}$ & plasma energy & 1.9761 & eV \\
$\gamma_\mathrm{D}$ & Drude damping & 0.0038 & eV \\
$A_\mathrm{A}$ & A-exciton oscillator strength & 0.2536 & -- \\
$E_{\mathrm{ex,A}}$ & A-exciton resonance energy & 1.9719 & eV \\
$\gamma_{\mathrm{ex,A}}$ & A-exciton damping & 0.0478 & eV \\
$A_\mathrm{B}$ & B-exciton oscillator strength & 0.0612 & -- \\
$E_{\mathrm{ex,B}}$ & B-exciton resonance energy & 2.3798 & eV \\
$\gamma_{\mathrm{ex,B}}$ & B-exciton damping & 0.06 & eV \\
$A_\mathrm{C}$ & C-exciton oscillator strength & 2.9692 & -- \\
$E_{\mathrm{ex,C}}$ & C-exciton resonance energy & 2.7619 & eV \\
$\gamma_{\mathrm{ex,C}}$ & C-exciton damping & 0.4036 & eV \\
$A_\mathrm{UV}$ & UV oscillator strength & 8.0269 & -- \\
$E_\mathrm{UV}$ & UV resonance energy & 4.2 & eV \\
$\gamma_\mathrm{UV}$ & UV damping & 2.0708 & eV \\
\hline
\end{tabular}
\end{table}

Comparing Figures~\ref{fig:Figure_5}(b) and (d) reveals the influence of the gold cladding on the guided optical modes. For the structure with $\SI{50}{\nano\meter}$ thick gold layers in Figure~\ref{fig:Figure_5}(b), one guided mode is visible and extends into the light cone. It can therefore contribute to the radiatively accessible dispersion. In contrast, for the structure with only $\SI{1}{\nano\meter}$ thick gold layers in Figure~\ref{fig:Figure_5}(d), the guided modes remain outside the air light cone, between the air and WS$_2$ light lines. This behavior is characteristic of guided modes in a dielectric slab waveguide. The comparison therefore shows that the gold layers modify the optical mode dispersion and allow the guided mode to appear inside the light cone.

The number of supported guided modes is also affected by the metallic cladding. While only one guided mode is visible for the full Au/WS$_2$/Au structure in Figure~\ref{fig:Figure_5}(b), two guided modes are present in the almost uncladded structure in Figure~\ref{fig:Figure_5}(d). Both of these modes cross the A-exciton energy and can therefore strongly couple to the excitonic resonance.
Using the uncoupled optical-mode energies extracted from Figures~\ref{fig:Figure_5}(b) and (d), together with the A-exciton energy of $\SI{1.97}{\electronvolt}$, Equation~\eqref{eq:polariton_energies} is used to calculate the lower and upper polariton branches shown in Figures~\ref{fig:Figure_5}(a) and (c). 

In addition to the guided modes, the same model is also fitted to the bent WS$_2$ light line, which also exhibits strong-coupling behavior in the calculated dispersions.
In Figure~\ref{fig:Figure_5}(a), the black lower- and upper-polariton branches correspond to the bent WS$_2$ light line, while the orange branches correspond to the guided mode. In both cases, a coupling strength of $G=\SI{190}{\milli\electronvolt}$ is obtained. In Figure~\ref{fig:Figure_5}(c), the black and orange polariton branches again correspond to the bent WS$_2$ light line and the first guided mode, respectively, while the additional pink branches arise from the second guided mode. The corresponding coupling strengths are $G_{\mathrm{black}}=\SI{190}{\milli\electronvolt}$, $G_{\mathrm{orange}}=\SI{180}{\milli\electronvolt}$, and $G_{\mathrm{pink}}=\SI{120}{\milli\electronvolt}$.
The coupling strength associated with the bent WS$_2$ light line remains unchanged when the gold layers are removed. In contrast, the coupling strength of the guided mode is reduced from $\SI{190}{\milli\electronvolt}$ in the full Au/WS$_2$/Au structure to $\SI{180}{\milli\electronvolt}$ in the almost uncladded structure. This indicates that the gold layers enhance the coupling between the guided mode and the WS$_2$ A-exciton by increasing the optical field confinement.

\FloatBarrier

\section{Conclusions}
\label{sec:conclusions}

In this work, we investigated Au/WS$_2$/Au heterostructures experimentally by angle- and spatially resolved cathodoluminescence spectroscopy and theoretically by calculating the optical mode dispersion from Maxwell's equations. The energy-momentum CL maps recorded for different WS$_2$ flake thicknesses show a pronounced thickness-dependent resonance below the A-exciton energy. For thinner flakes, this resonance approaches the A-exciton and exhibits a clear anticrossing behavior, consistent with the formation of guided exciton-polariton modes.
Additional vertical features observed in several angle-resolved CL spectra were assigned to interference between transition radiation and surface plasmon polaritons. This assignment is supported by the comparison with calculated interference maxima and by the spatially resolved CL spectrum, which shows both the guided-mode resonance and pronounced interference maxima. 

The spatially resolved measurement therefore confirms the local excitation of the guided lower-polariton resonance and provides additional evidence for TR--SPP interference in the Au/WS$_2$/Au structure.
To further identify the nature of the supported modes, we analyzed the calculated dispersion relations over an extended momentum range, including the region outside the light cone. This revealed that the resonance observed inside the light cone continues to larger in-plane momenta and forms a lower polariton branch with a pronounced anticrossing near the WS$_2$ A-exciton.
Finally, the coupling strength was extracted by fitting the calculated lower and upper polariton branches with a Hopfield model. For the full Au/WS$_2$/Au structure with $\SI{50}{\nano\meter}$ thick gold layers, a coupling strength of $G=\SI{190}{\milli\electronvolt}$ was obtained for the guided mode. In comparison, the corresponding almost uncladded WS$_2$ waveguide shows reduced coupling strengths of $G=\SI{180}{\milli\electronvolt}$ and $G=\SI{120}{\milli\electronvolt}$ for the supported guided modes. This comparison shows that the gold cladding modifies the guided-mode dispersion, brings the guided mode into the radiatively accessible momentum range, and enhances the exciton--photon coupling through increased optical field confinement.

Overall, the Au/WS$_2$/Au heterostructure provides a compact platform for guided exciton-polariton modes with enhanced coupling strength compared to related self-hybridized TMDC systems. In addition, the lower polariton branch exhibits a strongly reduced group velocity inside the light cone, making such metal-clad TMDC waveguides promising for future studies of slow polariton propagation and enhanced light--matter interaction.
These results may therefore provide a complementary route to microcavity-based polaritonic systems for future studies of polariton condensation, lasing, and optoelectronic applications.

\section*{Author Contributions}

F.M. performed the CL measurements, analyzed the data, and performed the simulations. M.T. performed the CL measurements and fabricated the sample. V.D.G. performed the AFM measurements. K.R. provided the WS$_2$ flakes. N.T. supervised the work. F.M. and N.T. wrote the manuscript. All co-authors contributed to discussions.

\section*{Acknowledgments}

This project received funding from the Volkswagen Foundation (Momentum Grant), the European Research Council (ERC Consolidator Grant UltraSpecT with No. 101170341; ERC Proof-of-Concept Grant UltraCoherentCL with No. 101157312), and Deutsche Forschungsgemeinschaft.

\section*{Conflicts of Interest}

The authors declare no conflicts of interest.

\section*{Data Availability}

The data that support the findings of this study are available from the corresponding author upon reasonable request.

\bibliographystyle{unsrtnat}
\bibliography{main}

\end{document}